# Some New Results on Binary Relations

Roy S. Freedman[1]

**Abstract**

It is well known that if a function from set *A* to set *B* has a right inverse then the function is a surjection and the right inverse is an injection. For finite sets, the number of functions, injections, and surjections can also be counted. Relations generalize functions: do similar results exist for relations? This paper proves several new results concerning binary relations. For finite sets, we derive formulas for the number of right total, right unique, left total, and left unique relations. We also provide formulas that count the number of relations that are both right unique and left unique; right unique and right total; and left unique and left total. We conclude by discussing the probability that a relation selected at random is right unique or right total.

## 1  Properties of Relations

Many sources on discrete mathematics [5,6], combinatorics [10], or functional programming [1,4] define and utilize the properties of relations. Denote sets by letters $A, B, C, \cdots$. Denote the cardinality of a set $A$ by $|A|$; thus for finite sets $A = \{a_1, a_2, \cdots, a_n\}$ and $B = \{b_1, b_2, \cdots, b_k\}$, $|A| = n$ and $|B| = k$. A relation $R$ between two sets $A$ and $B$ is any subset of the Cartesian product $A \times B$:

$$R \subseteq A \times B = \{(a,b) \mid a \in A \text{ and } b \in B\}.$$

We call set A the input set and set B the output set. Denote relations by capital letters $R, S, T, U, V, \cdots$.

We abbreviate the definition $R \subseteq A \times B$ by the relation specification $R : A \to B$. As in the case with function composition, a relation $R$ between sets $A$ and $B$ can be combined with another relation $S$ between set $B$ and set $C$. This relation creates a third relation $S \circ R : A \to C$ between sets $A$ and $C$ in the following way:

$$(a,b) \in R \text{ and } (b,c) \in S \text{ then } (a,c) \in S \circ R$$

Our notation for composition follows that for function composition: $S \circ R$ means first apply $R$ and then apply $S$. Relation composition is associative: $R \circ (S \circ T) = (R \circ S) \circ T$.

---

[1] R.S. Freedman is with Inductive Solutions, Inc., New York, and with the Department of Finance and Risk Engineering, NYU Polytechnic School of Engineering. Email: roy@inductive.com.

The identity relation on a set is defined by $1_A : A \to A$ with $1_A = \{(a,a) | a \in A\}$. The identity relation has the property that $1_B \circ R = R \circ 1_A = R$. It is the relation associated with a null action or transformation.

An inverse relation reverses the effects of the relation. Given relations $R : A \to B$, $S : B \to A$, and $T : B \to A$: if $R \circ S = 1_B$ and $T \circ R = 1_A$ then relation $S$ is called a right inverse and relation $T$ is called a left inverse. If the inverses both simultaneously exist then they are equal: associativity of relation composition implies:

$$T \circ 1_B = T \circ (R \circ S) = (T \circ R) \circ S = 1_A \circ S = S .$$

The opposite relation $R^o : B \to A$ is defined by switching the order of the pairs in $R$. So $R^o = \{(b,a) | (a,b) \in R\}$. One simple consequence of this definition is that $R \subseteq S$ if and only if $R^o \subseteq S^o$.

## 2 The Four Basic Intrinsic Properties of Relations

The property of having a right inverse or left inverse is *extrinsic*: the property depends on the existence of another relation. An *intrinsic* property of a relation is independent of other relations: it depends only on the relation itself. The four basic intrinsic properties are right unique, right total, left unique, and left total. Many of these properties are discussed in the work of Carnap [2, p. 125]; Suppes [8, p. 68]; and Pahl and Damrath [6, p. 506].

Being *right unique* implies that for any given input there is a single output: two outputs for the same input must be the same. More formally, a relation $R : A \to B$ is right unique if for $a \in A$ and $b_1, b_2, \in B$:

$$(a, b_1) \in R \text{ and } (a, b_2) \in R \text{ implies } b_1 = b_2 .$$

Denote the set of all right unique relations between sets $A$ and $B$ by $RU$.

One of the goals of relation calculus (also known as the "algebra of relations") is to create a formalism without relying on variables [3,9]. Such "point-free" representations result in properties and theorems that are easier to read and understand. For example, by examining the definitions of the opposite relation and the identity relation, we see that the definition of right unique is equivalent to

$$(a, b_1) \in R \text{ and } (b_2, a) \in R^o \text{ implies } (b_2, b_1) \in 1_B$$

By reviewing the definition of relation composition, this shows that a relation $R$ is right unique if the "point-free" formula $R \circ R^o \subseteq 1_B$ holds for relation $R$.

Being *right total* implies that every output is a result of some input: for each $b \in B$ there exists an $a \in A$ such that $(a, b) \in R$. This implies, for all $b \in B$

$$(a, b) \in R \text{ and } (b, a) \in R^o .$$

By reviewing the definition of relation composition, this implies that for every output $b \in B$, $(b,b) \in R \circ R^o$. Consequently, a relation $R$ is right total if $R \circ R^o \supseteq 1_B$. We denote the set of all right total relations between sets $A$ and $B$ by $RT$.

A relation $R$ is *left unique* if $R^o$ is right unique. This implies that, for $a_1, a_2, \in A$ and $b \in B$,

$$(a_1, b) \in R \text{ and } (a_2, b) \in R \text{ implies } a_1 = a_2.$$

Being left unique implies that for any given output there is a single input. A relation $R$ is left-unique if $R^o \circ R \subseteq 1_A$. Denote the set of all left unique relations between sets $A$ and $B$ by $LU$.

A relation $R$ is *left total* if $R^o$ is right total. This implies that every $a \in A$ has a corresponding output. A relation $R$ is left total if $R^o \circ R \supseteq 1_A$. Denote the set of all left total relations between sets $A$ and $B$ by $LT$.

A *function* is a relation that is left total and right unique. Denote functions by lower case letters $f, g, h, \cdots$. A function has a single input for a given output: thus, if $(a,b) \in f$, the unique (but arbitrary) output of a function given input $a$ is denoted by $b = f(a)$. The set of all functions between sets $A$ and $B$ is denoted by the intersection $RU \cap LT$. A *surjection* is a right total function: the set of all surjections between sets $A$ and $B$ is $RU \cap LT \cap RT$. An *injection* is a left unique function: the set of all injections between sets $A$ and $B$ is $RU \cap LT \cap LU$. A *bijection* is a function that is an injection and a surjection: the set of all bijections between sets $A$ and $B$ is $RU \cap LT \cap LU \cap RT$. Note that if $f$ and $f^o$ are both functions then $f$ and $f^o$ must be bijections. Note that the identity relation is also a bijection.

For illustration, the Appendix lists all relations $R : A \to B$ with $A = \{a1, a2, a3\}$ and $B = \{b1, b2\}$. The four intrinsic properties – showing whether a relation is a member of *RT, RU, LT,* or *LU* – are shown for each relation.

## 3 New Results Concerning Intrinsic and Extrinsic Properties

It is well known (and proved, e.g., in [5]) that if a function $f : A \to B$ has a right inverse $h : B \to A$ then function $f$ is a surjection and the right inverse $h$ is an injection:

$$f \circ h = 1_B \text{ implies } f \in RU \cap LT \cap RT \text{ and } h \in RU \cap LT \cap LU. \qquad (*)$$

This result links the intrinsic properties of functions (surjectivity, injectivity) with extrinsic properties (having right and left inverses). Relations generalize functions. In the following, we show several new results that generalize (*).

### 3.1 Right Inverses of Relations

We have the following generalization of the right inverse result (*) for functions:

**Theorem 1**. *Given relations $R : A \to B$, $S : B \to A$ and function $f : A \to B$.*

  *(1) If $R \circ S = 1_B$ then R is a right total relation and S is a left total relation.*

  *(2) If $f \circ S = 1_B$ then f is a surjection and S is both a left unique and left total relation.*

To prove this theorem, we utilize the following rules which were proved by Hoogendijk [4, p. 24] and Bird & de Moor, [1, p.90]:

**Lemma**. Given relations *R* and *S* and function *f*:

  Rule 1. $(R \circ S)^o = S^o \circ R^o$.

  Rule 2. $R \subseteq R \circ R^o \circ R$.

  Rule 3. $f \circ U \subseteq V$ if and only if $U \subseteq f^o \circ V$.

  $U \circ g^o \subseteq V$ if and only if $U \subseteq V \circ g$.

We now prove Theorem 1:

**Proof of (1)**.

Rule 2 in the Lemma, $R \subseteq R \circ R^o \circ R$ implies $R \circ S \subseteq (R \circ R^o \circ R) \circ S$. Applying the condition (1) in Theorem 1 implies $1_B \subseteq R \circ R^o$. Thus $R \in RT$. Also note that $R \circ S \subseteq R \circ S \circ S^o \circ S$ so $1_B \subseteq S^o \circ S$. Thus $S \in LT$. QED.

Note that the converse is not true. For example, consider $A = \{a_1, a_2\}$ and $B = \{b_1, b_2\}$. For $R^* = \{(a_1, b_1), (a_1, b_2), (a_2, b_1)\}$ and $S^* = \{(b_1, a_1), (b_2, a_2)\}$, then $R^*$ is right total and $S^*$ is left total. However, $R^* \circ S^* = \{(b_1, b_1), (b_1, b_2), (b_2, b_1)\} \neq 1_B$.

**Proof of (2)**.

First, note if $f \circ S = 1_B$ then $f \circ S \subseteq 1_B$. Now use Rule 3 in the Lemma. This implies, by replacing $U := S$ and $V := 1_B$ that $S \subseteq f^o$. Composition on both sides with $f$ implies $f \circ S = 1_B \subseteq f \circ f^o$. Thus $f \in RT$ so $f$ is a surjection. Rule 1 implies $S^o \circ S \subseteq S^o \circ f^o = (f \circ S)^o = 1_B$, so $S \in LU$. Now use the first part (1) of the condition so $S \in LT$ as well. Putting both together shows that $S \in LU \cap LT$. QED.

Note that that if *S* is also right unique in the condition of Theorem 1 part (2), then *S* is a function and the conclusion of the theorem is that $S \in RU \cap LU \cap LT$. This shows that *S* is an injection, so Theorem 1 part (2) also proves the right inverse result stated above as equation (*).

### 3.2   Closure Under Composition

The four intrinsic properties are closed under composition. In other words, the composition of two right unique relations is right unique; the composition of two right total relations is right total; the composition of two left unique relations is left unique; the composition of two left total relations is left total. More formally, we have the following:

**Theorem 2.** *Given relations* $R : A \to B$ *and* $S : B \to A$:
   (1) *If* $R, S \in RU$ *then* $R \circ S \in RU$.
   (2) *If* $R, S \in LU$ *then* $R \circ S \in LU$.
   (3) *If* $R, S \in RT$ *then* $R \circ S \in RT$.
   (4) *If* $R, S \in LT$ *then* $R \circ S \in LT$.

**Proof.**

*(1)* $(R \circ S) \circ (R \circ S)^o = (R \circ S) \circ (S^o \circ R^o) \subseteq R \circ 1_A \circ R^o \subseteq 1_B$.

*(2)* $(R \circ S)^o \circ (R \circ S) = (S^o \circ R^o) \circ (R \circ S) \subseteq S^o \circ 1_A \circ S \subseteq 1_B$.

*(3)* $(R \circ S) \circ (R \circ S)^o = (R \circ S) \circ (S^o \circ R^o) \supseteq R \circ 1_A \circ R^o \supseteq 1_B$.

*(4)* $(R \circ S)^o \circ (R \circ S) = (S^o \circ R^o) \circ (R \circ S) \supseteq S^o \circ 1_A \circ S \supseteq 1_B$. QED.

### 3.3 Closure Under Subset and Superset

We show that subsets of right unique relations are right unique and subsets of left unique relations are left unique. Moreover, a relation that is a superset of a right total relation is also right total, and a relation that is a superset of a left total relation is also left total.

**Theorem 3.** *Given relations* $R : A \to B$ *and* $S : A \to B$ *with* $R \subseteq S$.
(1) *If* $S \in RU$ *then* $R \in RU$.
(2) *If* $R \in RT$ *then* $S \in RT$.
(3) *If* $S \in LU$ *then* $R \in LU$.
(4) *If* $R \in LT$ *then* $S \in LT$.

**Proof.**
Note that $R \subseteq S$ implies that $R^o \subseteq S^o$. These imply

$$R \circ R^o \subseteq R \circ S^o \subseteq S \circ S^o.$$

We also have

$$R^o \circ R \subseteq R^o \circ S \subseteq S^o \circ S.$$

The remainder is straightforward:
(1) $S \in RU$ implies $S \circ S^o \subseteq 1_B$; this implies $R \circ R^o \subseteq 1_B$ so $R \in RU$.
(2) $R \in RT$ implies $R \circ R^o \supseteq 1_B$; this implies $S \circ S^o \supseteq 1_B$ so $S \in RT$.
(3) $S \in LU$ implies $S^o \circ S \subseteq 1_A$; this implies $R^o \circ R \subseteq 1_A$ so $R \in LU$.
(4) $R \in LT$ implies $R^o \circ R \supseteq 1_A$; this implies $S^o \circ S \supseteq 1_A$ so $S \in LT$. QED.

## 4 Cardinalities

The cardinality of the set of all relations from $A$ to $B$ is $2^{n \cdot k}$, where $n$ is the number of inputs (the cardinality of input set $A$) and $k$ the number of outputs (the cardinality of the output set $B$). Surprisingly, the cardinality of the set of right unique relations $|RU|$ or the cardinality of the set of right total relations $|RT|$ is not found in the literature.

Results concerning the cardinalities of functions are well known [5,10]. For finite input and output sets, the number of functions from $A$ to $B$ is

$$|RU \cap LT| = k^n$$

The number of surjections from $A$ to $B$ is known to be

$$|RU \cap LT \cap RT| = \sigma(n,k),$$

where $\sigma(n,k) = 0$ for $n < k$; $\sigma(n,1) = 1$; $\sigma(n,n) = n!$ and, in general, for $n \geq k$:

$$\sigma(n,k) = k^n + \sum_{j=1}^{k-1} \binom{k}{j} \cdot (k-j)^n \cdot (-1)^j.$$

Note that $\sigma(n,k)$ is related to the Stirling number of the second kind $S(n,k)$ [7] where the number of surjections $\sigma(n,k) = k! \cdot S(n,k)$.

The number of injections from $A$ to $B$ is denoted by

$$\eta(n,k) = |RU \cap LT \cap LU|,$$

where [5,10]

$$\eta(n,k) = \begin{cases} k \cdot (k-1) \cdot \ldots \cdot (k-n+1), & k \geq n \\ 0, & n > k \end{cases}$$

Our main result is the following theorem.

**Theorem 4.** *Given the set of all relations* $R: A \to B$. *Then*

$$|RU| = (k+1)^n - 1.$$
$$|LU| = (n+1)^k - 1.$$
$$|LT| = (2^k - 1)^n.$$
$$|RT| = (2^n - 1)^k.$$
$$|RU \cap LU| = \sum_{j=1}^{k} \binom{n}{j} \cdot \eta(j,k).$$
$$|RU \cap RT| = \sum_{j=k}^{n} \binom{n}{j} \cdot \sigma(j,k).$$
$$|LU \cap LT| = \sum_{j=n}^{k} \binom{k}{j} \cdot \sigma(j,n).$$

The proof is in the following sections. The properties of $\sigma(n,k)$ and $\eta(n,k)$ imply that $|RU \cap RT| = 0$ for $n < k$ and $|LU \cap LT| = 0$ for $n > k$. The formula for $|RT \cap LT|$ is a problem for future research.

## 4.1 Enumeration of Right Unique Relations

We obtain the number of right unique relations $R: A \rightarrow B$ by summing over all combinations of right unique and left total relations, with the combination taken over all inputs ranging from 1 to $n$. A right unique and left total relation is a function. The number of functions between two finite sets is $|RU \cap LT| = k^n$. Thus, the number of right unique relations is:

$$|RU| = \sum_{j=1}^{n} \binom{n}{j} \cdot k^j = (k+1)^n - 1.$$

The last equality results from using the binomial theorem. By switching inputs to outputs, the total number of left unique relations is:

$$|LU| = (n+1)^k - 1$$

Figure 1 shows the cardinalities of $RU$ for relations having up to six inputs and outputs.

| RU | k=1 | 2 | 3 | 4 | 5 | 6 |
|---|---|---|---|---|---|---|
| n=1 | 1 | 2 | 3 | 4 | 5 | 6 |
| 2 | 3 | 8 | 15 | 24 | 35 | 48 |
| 3 | 7 | 26 | 63 | 124 | 215 | 342 |
| 4 | 15 | 80 | 255 | 624 | 1295 | 2400 |
| 5 | 31 | 242 | 1023 | 3124 | 7775 | 16806 |
| 6 | 63 | 728 | 4095 | 15624 | 46655 | 117648 |

Figure 1. $|RU|$: the total number of relations that are right unique.

## 4.2 Enumeration of Relations that are both Right Unique and Left Unique

We obtain the total number of right unique and left unique relations by summing over all combinations of right unique, left unique and left total relations, with the combination taken over all inputs ranging from 1 to $n$. A right unique, left unique and left total relation is an injection and the number of injections between two finite sets is $|RU \cap LT \cap LU| = \eta(n,k)$ so:

$$|RU \cap LU| = \sum_{j=1}^{n} \binom{n}{j} \cdot \eta(j,k).$$

Since $\eta(j,k) = 0$ for $j > k$, the upper limit in the sum can be restricted to $k$ so we have our result. Figure 2 shows $|RU \cap LU|$ for relations having up to six inputs and outputs.

| RULU | k=1 | 2 | 3 | 4 | 5 | 6 |
|---|---|---|---|---|---|---|
| n=1 | 1 | 2 | 3 | 4 | 5 | 6 |
| 2 | 2 | 6 | 12 | 20 | 30 | 42 |
| 3 | 3 | 12 | 33 | 72 | 135 | 228 |
| 4 | 4 | 20 | 72 | 208 | 500 | 1044 |
| 5 | 5 | 30 | 135 | 500 | 1545 | 4050 |
| 6 | 6 | 42 | 228 | 1044 | 4050 | 13326 |

Figure 2. $|RU \cap LU|$: the total number of relations that are both right unique and left unique.

### 4.3   Enumeration of Relations that are both Right Unique and Right Total

We obtain the total number of right unique and right total relations by summing over all combinations of right unique, right total and left total relations, with the combination taken over all inputs ranging from 1 to $n$. A right unique, right total and left total relation is a surjection and the number of surjections between two finite sets is $|RU \cap LT \cap RT| = \sigma(n,k)$, so:

$$|RU \cap RT| = \sum_{j=1}^{n} \binom{n}{j} \cdot \sigma(j,k).$$

Since $\sigma(j,k) = 0$ for $j < k$, the lower limit in the sum can be restricted to $k$ so we have our result. Figure 3 shows the cardinalities of $RU \cap RT$ for relations having up to six inputs and outputs.

| RURT | k=1 | 2 | 3 | 4 | 5 | 6 |
|---|---|---|---|---|---|---|
| n=1 | 1 | 0 | 0 | 0 | 0 | 0 |
| 2 | 3 | 2 | 0 | 0 | 0 | 0 |
| 3 | 7 | 12 | 6 | 0 | 0 | 0 |
| 4 | 15 | 50 | 60 | 24 | 0 | 0 |
| 5 | 31 | 180 | 390 | 360 | 120 | 0 |
| 6 | 63 | 602 | 2100 | 3360 | 2520 | 720 |

Figure 3.   $|RU \cap RT|$: the total number of relations that are both right unique and right total.

Note that for $n < k$, $|RU \cap RT| = 0$.

By switching inputs to outputs, the total number of relations that are left unique and left total is

$$|LU \cap LT| = \sum_{j=n}^{k} \binom{k}{j} \cdot \sigma(j,n).$$

### 4.4   Enumeration of Relations that are Left Total

We obtain the total number of right total relations by first enumerating the cardinality of the set of left total relations. Note first that the null relation is not left total. Observe that for input set $\{a_1\}$ there are $2^k - 1$ non-empty left total relations – the set of tuples with indices $j_1, j_2, \ldots, j_n = 1, 2, \ldots, k$:

$$\{(a_1, b_{j_1})\}, \{(a_1, b_{j_1}), (a_1, b_{j_2})\}, \ldots \{(a_1, b_{j_1}), (a_1, b_{j_2}), \ldots, (a_1, b_{j_k})\}.$$

For input set $\{a_2\}$ there are also $2^k - 1$ non-empty left total relations – the set of tuples with indices $h_1, h_2, \ldots, h_n = 1, 2, \ldots, k$:

$$\{(a_2, b_{h_1})\}, \{(a_2, b_{h_1}), (a_2, b_{h_2})\}, \ldots \{(a_2, b_{h_1}), (a_2, b_{h_2}), \ldots, (a_2, b_{h_k})\}.$$

Now combine the two independent inputs. Enumerating the number of left total relations on input set $\{a_1, a_2\}$ results in $(2^k - 1)^2$ non-empty left total relations:

$$\{(a_1, b_{j_1}), (a_2, b_{h_1})\}, \{(a_1, b_{j_1}), (a_2, b_{h_1}), (a_2, b_{h_2})\}, \ldots, \{(a_1, b_{j_1}), (a_2, b_{h_1}), (a_2, b_{h_2}), \ldots, (a_2, b_{h_k})\} \ldots$$

By induction, for $n$ unique inputs, the total number of left total relations is:

$$|LT| = (2^k - 1)^n.$$

Consequently, by switching inputs to outputs, the total number of right total relations is:

$$|RT| = (2^n - 1)^k.$$

Figure 4 shows the cardinalities of $LT$ for relations having up to six inputs and outputs.

| LT  | k=1 | 2   | 3      | 4        | 5         |
|-----|-----|-----|--------|----------|-----------|
| n=1 | 1   | 3   | 7      | 15       | 31        |
| 2   | 1   | 9   | 49     | 225      | 961       |
| 3   | 1   | 27  | 343    | 3375     | 29791     |
| 4   | 1   | 81  | 2401   | 50625    | 923521    |
| 5   | 1   | 243 | 16807  | 759375   | 28629151  |
| 6   | 1   | 729 | 117649 | 11390625 | 887503681 |

Figure 4. $|LT|$: the total number of relations that are left total.

## 5 Probabilities

What is the probability that a relation is right unique? If our sample space contains equally likely relations, we have the following probabilities:

$$\Pr[R \in RU] = \frac{(k+1)^n - 1}{2^{n \cdot k}}$$

Note that for either large $k$ or $n$

$$\Pr[R \in RU] \approx \left(\frac{k+1}{2^k}\right)^n \to 0.$$

This implies that right unique relations are rare for large input or output sets. Similarly, for either large $k$ or $n$, left unique relations are rare, since

$$\Pr[R \in LU] \approx \left(\frac{n+1}{2^n}\right)^k \to 0.$$

Figure 5 shows the probabilities of a relation being right unique for relations having up to six inputs and outputs (probabilities rounded to 4 decimal places).

| Pr[RU] | k=1 | 2 | 3 | 4 | 5 | 6 |
|---|---|---|---|---|---|---|
| n=1 | 0.5000 | 0.5000 | 0.3750 | 0.2500 | 0.1563 | 0.0938 |
| 2 | 0.7500 | 0.5000 | 0.2344 | 0.0938 | 0.0342 | 0.0117 |
| 3 | 0.8750 | 0.4063 | 0.1230 | 0.0303 | 0.0066 | 0.0013 |
| 4 | 0.9375 | 0.3125 | 0.0623 | 0.0095 | 0.0012 | 0.0001 |
| 5 | 0.9688 | 0.2363 | 0.0312 | 0.0030 | 0.0002 | 0.0000 |
| 6 | 0.9844 | 0.1777 | 0.0156 | 0.0009 | 0.0000 | 0.0000 |

Figure 5.    $\Pr[R \in RU]$: the probability that a relation is right unique.

For right total relations,

$$\Pr[R \in RT] = \frac{(2^n - 1)^k}{2^{n \cdot k}} \approx \left(1 - \frac{1}{2^n}\right)^k.$$

Figure 6 shows the probabilities of a relation being right unique for relations having up to six inputs and outputs (probabilities rounded to 4 decimal places).

| Pr[RT] | k=1 | 2 | 3 | 4 | 5 | 6 |
|---|---|---|---|---|---|---|
| n=1 | 0.5000 | 0.2500 | 0.1250 | 0.0625 | 0.0313 | 0.0156 |
| 2 | 0.7500 | 0.5625 | 0.4219 | 0.3164 | 0.2373 | 0.1780 |
| 3 | 0.8750 | 0.7656 | 0.6699 | 0.5862 | 0.5129 | 0.4488 |
| 4 | 0.9375 | 0.8789 | 0.8240 | 0.7725 | 0.7242 | 0.6789 |
| 5 | 0.9688 | 0.9385 | 0.9091 | 0.8807 | 0.8532 | 0.8266 |
| 6 | 0.9844 | 0.9690 | 0.9539 | 0.9389 | 0.9243 | 0.9098 |

Figure 6.    $\Pr[R \in RT]$: the probability that a relation is right total.

Let $k = r \cdot 2^n$ for some $r > 0$. So

$$\Pr[R \in RT] \approx e^{-r}.$$

For large $r$ with $k > 2^n$ $\Pr[R \in RT] \to 0$, but for small $r$ with $k < 2^n$, $\Pr[R \in RT] \to 1$. Thus for a large number of outputs relative to inputs, $k \gg n$, the likelihood of a relation being right total is small; for a large number of inputs relative to outputs, $n \gg k$, it is likely that a relation is right total. Similarly, for left total relations,

$$\Pr[R \in RT] = \frac{(2^k - 1)^n}{2^{n \cdot k}} \approx \left(1 - \frac{1}{2^k}\right)^n$$

Thus for a large number of outputs relative to inputs, $k \gg n$, the likelihood of a relation being left total is high; for a large number of inputs relative to outputs, $n \gg k$, the likelihood of a relation being left total is small.

Figure 7 shows the probabilities of a relation being both right unique and right total for relations having up to six inputs and outputs (probabilities rounded to 4 decimal places).

| Pr[RU∩RT] | k=1 | 2 | 3 | 4 | 5 | 6 |
|---|---|---|---|---|---|---|
| n=1 | 0.5000 | 0.0000 | 0.0000 | 0.0000 | 0.0000 | 0.0000 |
| 2 | 0.7500 | 0.1250 | 0.0000 | 0.0000 | 0.0000 | 0.0000 |
| 3 | 0.8750 | 0.1875 | 0.0117 | 0.0000 | 0.0000 | 0.0000 |
| 4 | 0.9375 | 0.1953 | 0.0146 | 0.0004 | 0.0000 | 0.0000 |
| 5 | 0.9688 | 0.1758 | 0.0119 | 0.0003 | 0.0000 | 0.0000 |
| 6 | 0.9844 | 0.1470 | 0.0080 | 0.0002 | 0.0000 | 0.0000 |

Figure 7.  $\Pr[R \in RU \cap RT]$: the probability that a relation is right unique and right total.

The rarity of the right unique relations dominates the prevalence of the right total relations. Lower and upper bounds on $|RU \cap RT|$ are

$$\sigma(n,k) \leq \sum_{j=k}^{n} \binom{n}{j} \cdot \sigma(j,k) \leq (k+1)^n - 1.$$

The lower bound is established by taking $j = n$ in the summation. For the upper bound, first note that since the number of surjections is always less than the number of functions, $\sigma(j,k) \leq k^j$. Since $\sigma(j,k) = 0$ for $j < k$, starting the summation at $j = 1$ establishes the right inequality. For large $n$ and fixed $k$ with $n > k$, this upper bound implies:

$$\Pr[R \in RU \cap RT] \leq \frac{(k+1)^n - 1}{2^{n \cdot k}} \approx \left(\frac{k+1}{2^k}\right)^n \to 0.$$

This implies that relations that are both right unique and right total are rare for large input or output sets.

Appendix: Intrinsic properties: all relations $R: A \to B$ with $A = \{a1, a2, a3\}$ and $B = \{b1, b2\}$.

| Relation | n=3; k=2 | RT | LT | RU | LU |
|---|---|---|---|---|---|
| 1 | (a1, b1) | | | RU | LU |
| 2 | (a1, b2) | | | RU | LU |
| 3 | (a2, b1) | | | RU | LU |
| 4 | (a2, b2) | | | RU | LU |
| 5 | (a3, b1) | | | RU | LU |
| 6 | (a3, b2) | | | RU | LU |
| 7 | (a1, b1), (a1, b2) | RT | | | LU |
| 8 | (a1, b1), (a2, b1) | | | RU | |
| 9 | (a1, b1), (a2, b2) | RT | | RU | LU |
| 10 | (a1, b1), (a3, b1) | | | RU | |
| 11 | (a1, b1), (a3, b2) | RT | | RU | LU |
| 12 | (a1, b2), (a2, b1) | RT | | RU | LU |
| 13 | (a1, b2), (a2, b2) | | | RU | |
| 14 | (a1, b2), (a3, b1) | RT | | RU | LU |
| 15 | (a1, b2), (a3, b2) | | | RU | |
| 16 | (a2, b1), (a2, b2) | RT | | | LU |
| 17 | (a2, b1), (a3, b1) | | | RU | |
| 18 | (a2, b1), (a3, b2) | RT | | RU | LU |
| 19 | (a2, b2), (a3, b1) | RT | | RU | LU |
| 20 | (a2, b2), (a3, b2) | | | RU | |
| 21 | (a3, b1), (a3, b2) | RT | | | LU |
| 22 | (a1, b1), (a1, b2), (a2, b1) | RT | | | |
| 23 | (a1, b1), (a1, b2), (a2, b2) | RT | | | |
| 24 | (a1, b1), (a1, b2), (a3, b1) | RT | | | |
| 25 | (a1, b1), (a1, b2), (a3, b2) | RT | | | |
| 26 | (a1, b1), (a2, b1), (a2, b2) | RT | | | |
| 27 | (a1, b1), (a2, b1), (a3, b1) | | LT | RU | |
| 28 | (a1, b1), (a2, b1), (a3, b2) | RT | LT | RU | |
| 29 | (a1, b1), (a2, b2), (a3, b1) | RT | LT | RU | |
| 30 | (a1, b1), (a2, b2), (a3, b2) | RT | LT | RU | |
| 31 | (a1, b1), (a3, b1), (a3, b2) | RT | | | |
| 32 | (a1, b2), (a2, b1), (a2, b2) | RT | | | |
| 33 | (a1, b2), (a2, b1), (a3, b1) | RT | LT | RU | |
| 34 | (a1, b2), (a2, b1), (a3, b2) | RT | LT | RU | |
| 35 | (a1, b2), (a2, b2), (a3, b1) | RT | LT | RU | |
| 36 | (a1, b2), (a2, b2), (a3, b2) | | LT | RU | |
| 37 | (a1, b2), (a3, b1), (a3, b2) | RT | | | |
| 38 | (a2, b1), (a2, b2), (a3, b1) | RT | | | |
| 39 | (a2, b1), (a2, b2), (a3, b2) | RT | | | |
| 40 | (a2, b1), (a3, b1), (a3, b2) | RT | | | |
| 41 | (a2, b2), (a3, b1), (a3, b2) | RT | | | |
| 42 | (a1, b1), (a1, b2), (a2, b1), (a2, b2) | RT | | | |
| 43 | (a1, b1), (a1, b2), (a2, b1), (a3, b1) | RT | LT | | |
| 44 | (a1, b1), (a1, b2), (a2, b1), (a3, b2) | RT | LT | | |
| 45 | (a1, b1), (a1, b2), (a2, b2), (a3, b1) | RT | LT | | |
| 46 | (a1, b1), (a1, b2), (a2, b2), (a3, b2) | RT | LT | | |
| 47 | (a1, b1), (a1, b2), (a3, b1), (a3, b2) | RT | | | |
| 48 | (a1, b1), (a2, b1), (a2, b2), (a3, b1) | RT | LT | | |
| 49 | (a1, b1), (a2, b1), (a2, b2), (a3, b2) | RT | LT | | |
| 50 | (a1, b1), (a2, b1), (a3, b1), (a3, b2) | RT | LT | | |
| 51 | (a1, b1), (a2, b2), (a3, b1), (a3, b2) | RT | LT | | |
| 52 | (a1, b2), (a2, b1), (a2, b2), (a3, b1) | RT | LT | | |
| 53 | (a1, b2), (a2, b1), (a2, b2), (a3, b2) | RT | LT | | |
| 54 | (a1, b2), (a2, b1), (a3, b1), (a3, b2) | RT | LT | | |
| 55 | (a1, b2), (a2, b2), (a3, b1), (a3, b2) | RT | LT | | |
| 56 | (a2, b1), (a2, b2), (a3, b1), (a3, b2) | RT | | | |
| 57 | (a1, b1), (a1, b2), (a2, b1), (a2, b2), (a3, b1) | RT | LT | | |
| 58 | (a1, b1), (a1, b2), (a2, b1), (a2, b2), (a3, b2) | RT | LT | | |
| 59 | (a1, b1), (a1, b2), (a2, b1), (a3, b1), (a3, b2) | RT | LT | | |
| 60 | (a1, b1), (a1, b2), (a2, b2), (a3, b1), (a3, b2) | RT | LT | | |
| 61 | (a1, b1), (a2, b1), (a2, b2), (a3, b1), (a3, b2) | RT | LT | | |
| 62 | (a1, b2), (a2, b1), (a2, b2), (a3, b1), (a3, b2) | RT | LT | | |
| 63 | (a1, b1), (a1, b2), (a2, b1), (a2, b2), (a3, b1), (a3, b2) | RT | LT | | |
| | Totals: | 49 | 27 | 26 | 15 |